\documentclass[10pt,conference]{IEEEtran}

\usepackage[english]{babel}
\usepackage[utf8]{inputenc}
\usepackage[]{todonotes}
\usepackage{hyperref}
\usepackage{url}
\usepackage{booktabs}
\usepackage{lscape}
\usepackage{longtable}
\usepackage{tabularx}
\usepackage{xspace}
\usepackage{multirow}
\usepackage{float}
\usepackage{array}
\usepackage{graphicx}
\usepackage{pbox}
\usepackage{makecell}
\usepackage{color}
\usepackage{colortbl}
\usepackage{csvsimple}
\usepackage[nottoc]{tocbibind}
\usepackage{framed}
\usepackage{xcolor}

\usepackage{flushend}

\usepackage{enumitem}
\setitemize{noitemsep,topsep=0pt,parsep=0pt,partopsep=0pt}
\setlist[itemize]{leftmargin=*}

\newcounter{problem}
\newcommand{\problem}[1]{\stepcounter{problem}\color{teal}\textbf{#1 (P\theproblem)}\color{black}\xspace}
\newcommand{\problemref}[1]{\color{teal}\textit{(#1)}\color{black}\xspace}

\newcommand{\lesson}[1]{\color{teal}\textbf{\textit{#1}}\color{black}\xspace}

\definecolor{benefitsbg}{HTML}{214517}
\definecolor{drawbacksbg}{HTML}{c62f2d}
\definecolor{poibg}{HTML}{33ddff} %

\colorlet{benefitsfg}{white}
\colorlet{drawbacksfg}{white}
\colorlet{poifg}{black}

\newcommand{\beginmrtable}{
	\begin{table}[H]
		\centering \begin{tabular}{|p{0.1cm}|p{0.9\linewidth}|}
		\hline  
}			

\newcommand{\addtbrow}[2]{
	\cellcolor{#1} & #2 \\ \cline{2-2}
}

\newcommand{\addlasttbrow}[5]{
	\multirow{-2}{*}[#1]{\cellcolor{#2} \rotatebox[origin=c]{90}{\color{#3}\textbf{#4}}} & #5 \\ \cline{2-2}
}

\newcommand{\closemrtable}{
	\end{tabular}\end{table}
}

\title{System- and Software-level Architecting Harmonization Practices for Systems-of-Systems\\ {\Large An exploratory case study on a long-running large-scale scientific instrument}} 

\author{
	\IEEEauthorblockN{H\'ector Cadavid\IEEEauthorrefmark{1}\IEEEauthorrefmark{4}, Vasilios Andrikopoulos\IEEEauthorrefmark{1}, Paris Avgeriou\IEEEauthorrefmark{1}, P. Chris Broekema\IEEEauthorrefmark{2}\IEEEauthorrefmark{3}			\\\{h.f.cadavid.rengifo, v.andrikopoulos, p.avgeriou\}@rug.nl, broekema@astron.nl
		\IEEEauthorblockA{\IEEEauthorrefmark{1}University of Groningen, Groningen, The Netherlands}
		\IEEEauthorblockA{\IEEEauthorrefmark{2}Netherlands Institute for Radio Astronomy (ASTRON), Dwingeloo, the Netherlands}
		\IEEEauthorblockA{\IEEEauthorrefmark{3}University of Cambridge, United Kingdom of Great Britain and Northern Ireland}
		\IEEEauthorblockA{\IEEEauthorrefmark{4}Escuela Colombiana de Ingenier\'ia, Bogot\'a, Colombia}
	}
}

\begin{document}

\IEEEoverridecommandlockouts
\IEEEpubid{\begin{minipage}{\textwidth}\ \\[12pt]
		Accepted for publication at ICSA 2021. \copyright 2021 IEEE
\end{minipage}} 

	\maketitle		
    \begin{abstract}
    The problems caused by the gap between system- and software-level architecting practices, especially in the context of Systems of Systems where the two disciplines inexorably meet, is a well known issue with a disappointingly low amount of works in the literature dedicated to it.
    At the same time, organizations working on Systems of Systems have been developing solutions for closing this gap for many years now.
    This work aims to extract such knowledge from practitioners by studying the case of a large-scale scientific instrument, a geographically distributed radio telescope to be more specific, developed as a sequence of projects during the last two decades.
    As the means for collecting data for this study we combine online interviews with a virtual focus group of practitioners from the organization responsible for building the instrument.   Through this process, we identify persisting problems and the best practices that have been developed to deal with them, together with the perceived benefits and drawbacks of applying the latter in practice.
    Some of our major findings include the need to avoid over-reliance on the flexibility of software to compensate for incomplete requirements, hidden assumptions, as well as late involvement of system architecting, and to facilitate the cooperation between the involved disciplines through dedicated architecting roles and the adoption of unifying practices and standards.    
    \end{abstract}

    \begin{IEEEkeywords}
    systems of systems, architecting, case study, scientific instruments, empirical software engineering
    \end{IEEEkeywords}	

	\section{Introduction}

 \textit{Systems of Systems} (SoS) are comprised of independent systems that cooperate to provide new capabilities~\cite{ISO21839}, particularly in industries such as defense, automotive, energy, and health care~\cite{pyster2015exploring}. In the development of SoS, the Systems Engineering (SE) and Software Engineering (SWE) disciplines have become highly interdependent: the development of a whole SoS is usually governed by a SE process given its scale and the number of disciplines it involves~\cite{muller2012validation,sheard2019systems}; at the same time, however, software is nowadays not only pervasive in most SoS, but also the predominant element of their offered features and qualities~\cite{pyster2015exploring,fairley2019systems}.

The combination of these two disciplines during the architecting activity remains a challenge, not only for SoS, but for most complex engineered systems in general. Maier~\cite{maier2006system} has pointed out that SE and SWE architecting and design approaches are often difficult to align when applied together. Sheard et al.~\cite{sheard2019systems} reported that they often interfere with each other's practices. For instance, Systems Engineers have traditionally followed a top-down, hierarchical functional decomposition of the envisioned system~\cite{fairley2019systems}. On the other hand, Software Engineers usually decompose software systems following architectural styles (e.g., \textit{Layers}, \textit{Pipes and Filters}, etc.) that do not enforce such hierarchical system/subsystem structure, or even prescribe a non-hierarchical one (e.g., \textit{Event-Driven}). Related work in this field~\cite{crowder2015multidisciplinary,gagliardi2009uniform,boehm2010extending}, and recent practitioner surveys~\cite{cadavid2020survey,muscarella2020systems} suggest that the gaps and mismatches between system-level and software-level architectures created by these interdisciplinary differences are linked to major \emph{integration and operation problems} in many engineered systems, including SoS.

Due to these challenges, several approaches and standards have been proposed for the harmonization of the practices of the two disciplines. A prominent example is the ISO~15288~\cite{ISO15288} standard for SE processes, which includes suggestions for its integration with ISO~12207-compliant~\cite{singh1996international} SWE processes. However, practices for harmonizing system- and software-level \emph{architecting} processes, in the context of SoS, have not been sufficiently explored in the literature. This holds true even in the case of SoS-tailored methodologies such as the \textit{Systems Engineering Guide for SoS}~\cite{DoD2008SoSSE}, or the \textit{DANSE}~\cite{danse}, \textit{COMPASS}~\cite{compass} and \textit{AMADEOS}~\cite{amadeos} guidelines. These all lack elements for integrating SWE and SE architecting practices.

Despite this lack of guidance, the significant number of existing operational SoS in the ``real world'' suggest that practitioners, have managed  to deal ---to some extent--- with the challenges of combining system- and software-level architecting practices. Therefore, it is likely they have discovered, over the years, valuable harmonization practices through trial-and-error. In this study, we aim to systematically \textit{identify and characterize such practices} in the domain of SoS.

To this end, we conducted a case study at ASTRON\footnote{\url{https://www.astron.nl}}, the Netherlands Institute for Radio Astronomy. We focus on a particular type of complex SoS, that of \textit{large-scale scientific instruments}.
ASTRON is ideal for this purpose because as an organization it has built and is operating a number of radio telescopes, the most recent being LOFAR~\cite{Haarlem:2013}, and currently contributes to the Square Kilometer Array (SKA)\footnote{\url{https://www.skatelescope.org/the-ska-project/}} project which aims to build a massive new radio telescope in South Africa and Western Australia. As case subjects, we analyzed the initial LOFAR project, and its follow-up, LOFAR2.0. These two projects together represent over 20 years of experience in the architecting, design and development of radio astronomy instruments.

The rest of this paper is structured as follows: Section~\ref{sec:related} summarizes the works related to this study. Section~\ref{sec:design} presents the study design, and Section~\ref{sec:results} elaborates on the results. Section~\ref{sec:discussion} discusses the relevance of our findings for the domain of SoS architecting in general, and Section~\ref{sec:lessons} summarizes the lessons we learned while conducting this study during the months of the COVID-19 pandemic. In Section~\ref{sec:ttv} we discuss threats to the validity of our work. Finally, Section~\ref{sec:conclusion} concludes the study.

	\section{Related work}
	\label{sec:related}
		
		To the best of our knowledge, only two studies so far have explored concrete practices for harmonizing system- and software-level architecting processes in SoS. 
More specifically, Boehm et al.~\cite{boehm2007using}, on their work on the Integrated Commitment Model, proposed an approach to address operational problems in SoS (e.g., service delays, conflicting plans) caused by the traditional hardware-centered systems engineering and acquisition practices of Consistutent Systems. Gagliardy et al.~\cite{gagliardi2009uniform}, on the other hand, proposed an evaluation method for SoS and software architectures ---early in the development process--- to address the lack of attention on the system quality attributes caused by the diversity of notations used for system and software elements of SoS.

Beyond these studies in the context of SoS, there is more related work that explores problems and practices between these two disciplines in systems in general. First, Maier's seminal paper on System and Software Architecture reconciliation \cite{maier2006system}, is one of the earliest studies on how to harmonize SE and SWE practices from an architectural perspective. This work, which later would become part of \textit{The art of systems architecting} book \cite{maier2009art}, pointed out the problems caused by gaps and mismatches between the traditional, hardware-centered systems engineering architectural structures and the modern software engineering ones. When it comes to empirical research in SE and SWE interplay, two studies supported by INCOSE\footnote{International Council on Systems Engineering} are among the earliest in this category.  First, Pyster et al.~\cite{pyster2015exploring} explored the problems between these two disciplines through a workshop with practitioners, whose results were late integrated into the \textit{Systems Engineering Body of Knowledge}\footnote{\url{https://www.sebokwiki.org/}} (SEBoK). A few years later, a study supported by INCOSE's \textit{Systems and Software Interface Working group} further explored these interdisciplinary problems and a series of  `best-practices' ---some of them related to architecting practices--- through a series of interviews with Systems Engineers and Software Engineers from different domains~\cite{muscarella2020systems}.

	\section{Study Design}
	\label{sec:design}
	
	\subsection{Research setting}

	ASTRON is a world-leading scientific institute dedicated to the development and scientific exploitation of radio telescopes. Radio telescopes, in contrast to optical telescopes, sample much lower frequencies in the electromagnetic spectrum. As a consequence, they need to be much larger to attain usable resolutions. Hence, modern radio telescopes, like LOFAR~\cite{Haarlem:2013}, consist of multiple, geographically distributed receivers or sets of receivers that are combined into a single virtual receiver in software. LOFAR, and more generally any large-scale distributed radio telescope, can thus be seen as a collection of independent but coherent collectors that sample, digitize, filter and transport the same electromagnetic wave front to be processed at a central location into scientific data products.

	In this sense, LOFAR's constituents  are stations with clusters of antennas that exhibit operational and managerial independence; however at the same time they are directed towards achieving a centrally managed purpose (e.g., large-sky surveys). LOFAR therefore, according to the categories of SoS identified by Maier~\cite{maier1998architecting}, can be classified as a \textit{directed SoS}. More specifically, given the profiling model proposed by Firesmith~\cite{firesmith2010profiling}, it could be described as an \textit{ultra-large-scale} \textit{directed SoS} with a \textit{high level of complexity}, made out of \textit{globally distributed}, \textit{independently governed} and \textit{operationally independent} systems. 
	The development of an SoS with such characteristics, in such a long time frame (over a dozen years), involved numerous unprecedented challenges, which led to a plethora of lessons learned for future projects. LOFAR2.0, an ongoing expansion of the scientific and technical capabilities of LOFAR (expected to be ready in 2025), is a follow-up project, whose development relies on the experience gathered from its predecessor~\cite{Hessels:2016}.

	\subsection{Objectives and Research Questions}\label{sec:obj-n-rrqq}
	
	For a precise definition of the goal of this study, we use the goal template proposed by the Goal-Question-Metric (GQM)~\cite{basili1992software} as follows: %
	
	\begin{framed}
	\begin{quote}
	\textbf{Analyze} the architecting process of systems with SoS characteristics
	\textbf{for the purpose} of identifying and characterizing specific harmonization practices
	\textbf{with respect to} system- and software-level architecting processes
	\textbf{from the viewpoint of} System and Software architects
	\textbf{in the context of} large-scale, multidisciplinary projects developing scientific instruments.
	\end{quote}
	\end{framed}
	
	Given this goal, and the research settings discussed above, the following research questions are addressed by this study:
	
	\begin{framed}
	\begin{description}

		\item[\textit{RQ1}] What problems, caused by  gaps or mismatches between system- and software-level architecting activities in SoS, have been identified by practitioners?

		\item[\textit{RQ2}] What specific practices have been developed or adopted by practitioners, %
		to harmonize architecting processes from Systems Engineering and Software Engineering?
		\item[\textit{RQ3}] What are the benefits and drawbacks of such developed/adopted practices, and how can they be further improved?

	\end{description}
	\end{framed}

	The first research question aims to identify problems, mainly of integration and operational nature as we know from the literature~\cite{crowder2015multidisciplinary,gagliardi2009uniform}, that appeared due to gaps and mismatches between system- and software-level practices in the architecting process of LOFAR (only). Gaps refer to missing elements on either side, while mismatches refer to discrepancies or incompatibilities between existing elements. The answer to this research question is expected to provide insights in the domain of SoS by exploring gaps/mismatches related to the SoS-related characteristics that, according to Firesmith's profiling model~\cite{firesmith2010profiling}, are exhibited to a higher degree in LOFAR, namely \textit{managerial independence}, \textit{size}, \textit{complexity} and \textit{physical distribution}.
	The second research question seeks to identify how those gaps or mismatches between system/software architecting processes have been addressed in LOFAR2.0 by adopting existing or developing new harmonization practices. Finally, the third research question aims to identify the pros and cons of said practices, alongside points for improvement, as perceived by the practitioners.

	\subsection{Research method}		
	
	Addressing the research questions proposed in this study requires achieving a deeper understanding of encountered problems, emerging solutions and lessons learned in real-life projects in the studied domain. 
	In other words, it entails investigating a phenomenon within its real-life context. We thus opt for a case study design as an empirical research method. More specifically, this study is designed as an exploratory, embedded multiple-case study~\cite{runeson2009guidelines}, where the LOFAR and LOFAR2.0 projects are the cases, and the architects that have been involved in them represent individual units of analysis. 
	
	\subsection{Data collection}		
	
\begin{figure}
	\centering
	\includegraphics[width=0.8\linewidth]{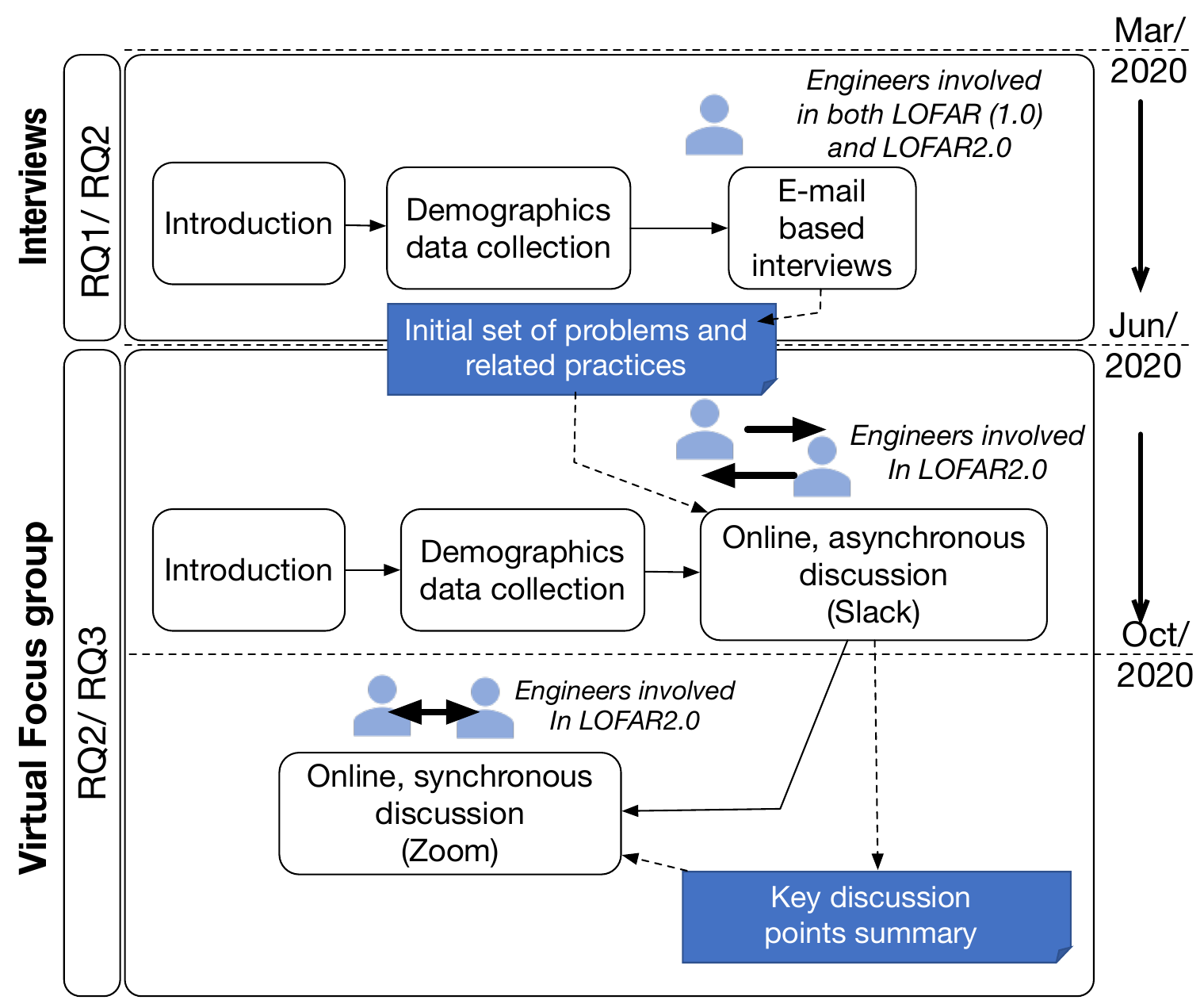}
	\caption{Description of the data collection approaches combined in the study. Solid arrows describe the activities sequence, and dashed ones the data carried between them. }
	\label{fig:data-collection}
\end{figure}

	Data collection took place from March 2020 to October 2020. This meant that in addition to the limited availability and geographical constraints of the participants that were a known concern when the study was designed, data collection was also constrained by the challenges posed by the \textit{COVID-19} pandemic. This even included a period of strict lockdown that forbade any in-person data collection.
	As a reaction to this situation, we adopted two qualitative data collection techniques in this case study: \textit{online asynchronous interviews} and \textit{virtual focus groups}\footnote{Both interview and focus group detailed guidelines are available at \url{https://doi.org/10.6084/m9.figshare.13332659.v1}]} as shown in Fig.~\ref{fig:data-collection}.
	The following sections elaborate on both techniques.

	\subsubsection{Online, text-based asynchronous interviews}\label{sec:text-based-interviews}

	Online, text-based asynchronous interviews, often referred to as email-based interviews,  are an alternative to the face-to-face interviews used extensively in domains such as health~\cite{hawkins2018practical,miner2012conducting} and social sciences~\cite{adnan2012computer}. Unlike traditional interviews, which are conducted face-to-face at a pre-arranged time, online/asynchronous interviews involve a similar conversation, but over an extended period of time by multiple email exchanges.
	Although face-to-face interviews are generally preferred over asynchronous, text-based interviews, as a data collection approach~\cite{bowden2015interviewing} the latter are particularly useful due to this study's restrictions, as discussed above. 
	Since participants can reply to questions at their own convenience, the time and resources required to participate is in principle minimized.

	The interviews followed a semi-structured approach, meaning that although there is a fixed set of questions, the interviews were open to follow-up, open-ended questions. In total, five architects involved in both the LOFAR and LOFAR2.0 projects, with an average of seven years of experience in LOFAR and three in LOFAR2.0, were interviewed. Two of them reported both SE and SWE as their main engineering practice areas, another two both SE and Electrical Engineering (EE), while the final one reported only SE as their engineering practice area within the project. Their respective academic background includes physics, electrical engineering, and astronomy. While none of these practitioners has a degree in SWE or computer science in general, they do have significant experience in those fields through their work for ASTRON and have received training on software architecture.%

	In the first part of the interview, and to address \textit{RQ1}, participants were asked about architecting scenarios in LOFAR where problems caused by interdisciplinary gaps or mismatches were perceived. In addition to the scenarios pointed out by the participants, we used related insights gathered from LOFAR's post-project review report \cite{ASTRON-internal-report} to further steer the conversation. Following up, participants were asked about architecting scenarios where the interdisciplinary gaps or mismatches were linked to the SoS-related characteristics of the LOFAR project. Once the provided scenarios and their related gaps/mismatches were sufficiently clarified, a new series of email exchanges focused on \textit{RQ2} by discussing the practices adopted, as a consequence of said problems, in LOFAR2.0. To ensure that the provided information was interpreted correctly by the interviewer, and that each new follow-up question has the right context for the interviewee, a summary of the points previously raised was provided on each email exchange. 
	The outcome of this process was a set of identified problems and the developed or adopted best practices to address them.

	\subsubsection{Virtual focus groups}

	The focus group technique is a qualitative research  methodology that aims to collect data through group interaction on a topic determined by the researcher~\cite{kontio2008focus}. In the focus group, unlike interviews, the information given by each participant can be viewed or even influenced by the others. A Virtual Focus Group (VFG), also known as an \textit{online focus groups}, is a variation of the traditional face-to-face focus group technique, where the discussions are mediated by a GSS (Group Support System) technology. 
	This method was selected to triangulate the findings of the interviews on \textit{RQ2}, and to address \textit{RQ3}. 
	
	Given that the focus group was centered on LOFAR2.0's actual architecting practices, the target population included engineers with architecting experience working on LOFAR 2.0, without necessarily them having experience on LOFAR. A total of ten people involved in the LOFAR2.0 project agreed to participate in this part of the study. They include one of the project managers, five software engineers, two systems engineers, one engineer on both practice areas (SE/SWE), and a system stakeholder and PI for some of the sub-projects under the umbrella of LOFAR2.0. The respective academic background of the participants included astronomy, computer science, electrical engineering and physics.%

	The seed discussion topics used in the VFG were based on the practices adopted in LOFAR2.0 (\textit{RQ2}) as identified by the online interviews in the previous step. In order to triangulate \textit{RQ2}-related results, and address \textit{RQ3}, we opted for a combination of asynchronous and synchronous VFG format using the popular services \textit{Slack} and \textit{Zoom}, respectively. The asynchronous part was intended to enable a discussion on \textit{RQ2} and \textit{RQ3} where participants could join at their convenience, giving more time to the moderators to  analyze the entries so they can create an appropriate discussion environment and encourage group discussion (e.g., regularly asking additional questions, clarifying participant's entries, etc.) \cite{stewart2005researching}. Following the advice of the most commonly cited VFG guidelines~\cite{stewart2005researching,murray1997using,moloney2003using}, only a few topics were delivered to the group at a time, aiming for a more in-depth discussion. 
	
	The second part of the VFG ---the synchronous one, through a Zoom call--- was also oriented towards \textit{RQ2} and \textit{RQ3} and elaborated on the points raised in the asynchronous part, facilitating a more focused discussion, which, in turn, was enriched by the socio-emotional cues that were previously absent. At the end of this meeting, and due to the perceived under-representation of the Systems Engineering point of view, an additional interview, based on the VFG findings, was conducted with one of the lead system engineers of LOFAR2.0, also through Zoom. The proposed guidelines for both synchronous and asynchronous parts of the VFG aimed to encourage participants to exchange views, experiences or anecdotes related to the harmonization practices identified. These could then be classified as benefits or drawbacks of applying said practices, or possible points of further improvement.

	\subsection{Data analysis}\label{sec:data-analysis}
		
The analysis of the e-mail-based interviews to address \textit{RQ1} was performed following an inductive \textit{Qualitative Content Analysis} (QCA)~\cite{elo2008qualitative}, also known as \textit{open coding}. Open coding involves an iterative process where codes are assigned to dataset samples, and then collapsed into a smaller number of higher-order categories. As this process is inductive, the code categories arise from the qualitative data ---in this case, categories of gaps and mismatches--- instead of using a predefined set of code categories. 

For the analysis of the VFG, and to address \textit{RQ2} and \textit{RQ3}, the same \textit{open coding} approach was followed to distill the high-order categories of practices adopted in LOFAR2.0. In this case, the qualitative data are extracted from both the text-based discussions and from the transcripts of the synchronous part of the  VFG. Furthermore, \textit{axial coding}, where the codes are related to each other through a combination of inductive and deductive thinking, is performed. More specifically, it is focused on identifying relations between the practices and the gaps/mismatches identified for \textit{RQ1}. Both \textit{open} and \textit{axial} coding were supported by the Atlas.ti\footnote{See https://atlasti.com} QCA software tool.

	\section{Results}
	\label{sec:results}

\begin{figure}
	\centering
	\includegraphics[width=0.9\linewidth]{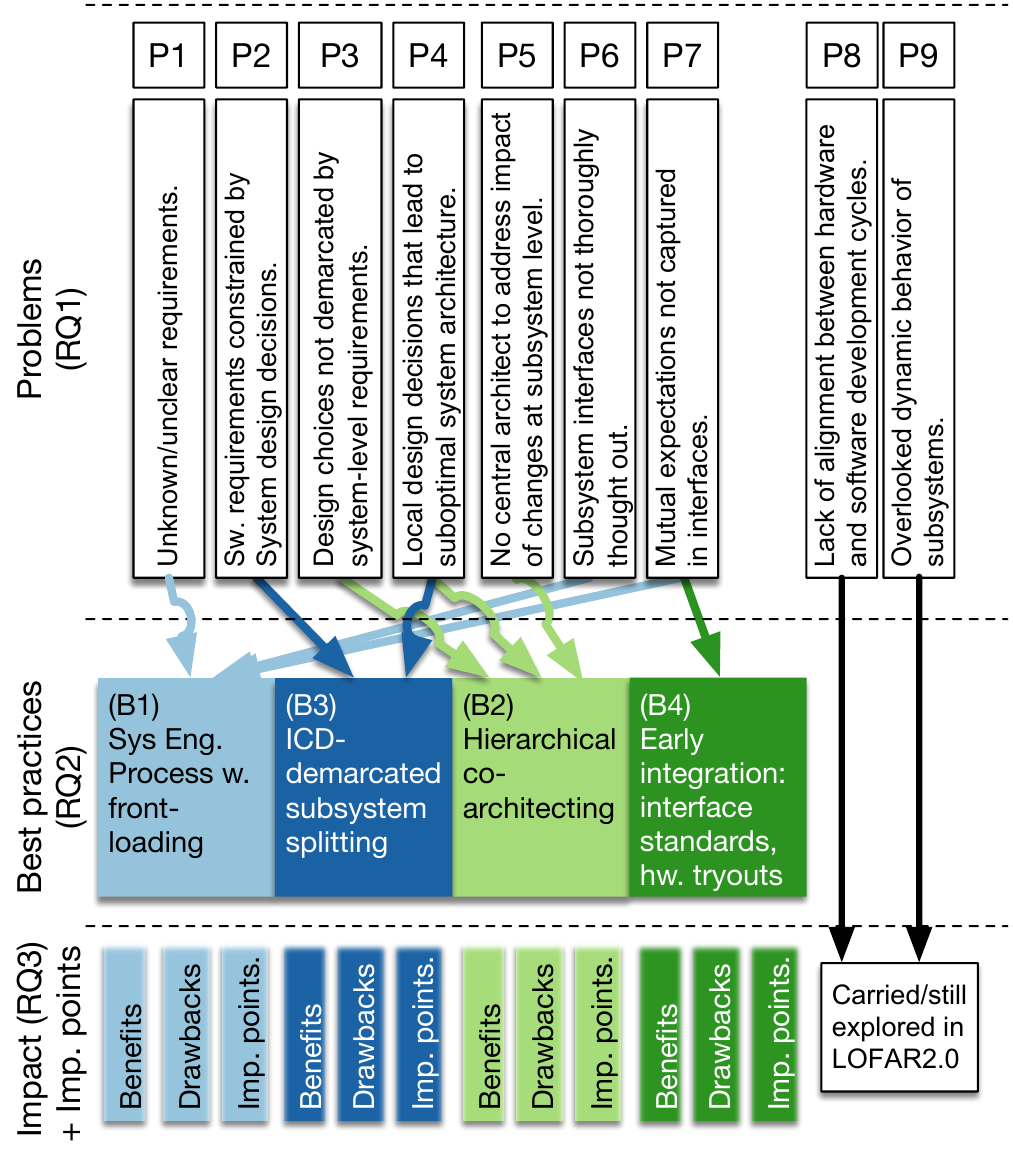}
	\caption{Problems, best practices, their perceived impact (benefits and drawbacks), and points of improvement, related to gaps and mismatches between the system-level and software-level architecting processes.}
	\label{fig:problems-practices-mapping}
\end{figure}

The following subsections describe the findings of this study, summarized graphically in Fig.~\ref{fig:problems-practices-mapping}. 
First, the problems caused by the gaps and mismatches between system- and software-level architecting activities (\textit{RQ1})
are described. Then, the practices that have been adopted to harmonize the architecting processes (\textit{RQ2}) are discussed, including their perceived impact on LOFAR2.0 (benefits and drawbacks), and the suggested points of improvement (\textit{RQ3}).

\subsection{Problems caused by system- and software-level architecting activities gaps and mismatches (\textit{RQ1})}

\textit{RQ1} as described in Section \ref{sec:obj-n-rrqq}, aims at identifying integration or operational problems in LOFAR, due to gaps or mismatches in system- and software-level architecting activities. It is worth noting that some of the activities and problems  described below refer to \textit{subsystem-level architecture} instead of \textit{software-level architecture}. In these cases, the problem described applies not only to pure software subsystems, but also to hardware- or software-intensive subsystems. The identified problems are classified, in turn, using three of the five activities of the architecture design process as proposed by Hofmeister et al.~\cite{hofmeister2005generalizing} and extended by Tang et al.~\cite{tang2010comparative}:

\begin{itemize}
	\item \textit{Architectural Analysis}:  where the architects come up with architecturally significant requirements based on the architectural concerns.
	\item \textit{Architectural Synthesis}: where the architects design a solution based on the architecturally significant requirements identified in the Analysis activity.
	\item \textit{Architectural Implementation}: where the architecture is realized by designers (i.e., a detailed design).
\end{itemize}

It is worth noting that the activity of \textit{Architectural Evaluation}, which gauges the quality of architectural decisions, is not included as none of the identified problems are related to it. Arguably, this could be due to the fact that the \textit{Architectural Evaluation} on the study's cases is done through the delivery of the instrument itself (that is, the radio telescope), which seems to be a common practice in SoS~\cite{cadavid2020survey}. In the following, we discuss these three categories of problems.

\subsubsection{Analysis-related problems}\label{sec:analysis-probs}

 In the email-based interviews it was pointed out that LOFAR was originally proposed as an instrument concept taking advantage of a technological opportunity, not as an instrument designed to answer a specific scientific question. Stakeholders were asked what they could do with the capabilities provided, rather than what they needed for their research. Furthermore, the subjects indicated that, in the system architecting process, the guidance on the software architecture was much missed. For these reasons, \problem{requirements for the software part of the project were, in many cases, unknown or not clear}, particularly regarding what type of control, monitoring, and scheduling was needed. One of the consequences for the software teams was that it was not clear to them what needed to be verified exactly, and hence tests were defined on-the-fly instead of being designed in advance.
 Furthermore, once the software development started, many design choices had already been made (e.g., due to hardware being in an advanced state of development), making the \problem{software requirements also severely constrained by design decisions at system-level}.

\subsubsection{Synthesis-related problems}\label{prob:synthesis}

According to the interviewees, in LOFAR there was little application of SE techniques, and the instrument was mostly seen as a collection of interrelated systems rather than a large system, each one being developed by a different group. This, and the requirements issues described above were the reasons that \problem{the room for design choices on the subsystems was not clearly demarcated by properly defined system-level requirements}; as a result, subsystem-level design choices were not properly mapped to system-level requirements. This led to situations where the benefits of local optimizations were exceeded by the problems caused by them in the overall project (e.g., design decisions that simplify firmware development having a large negative impact on the software that uses it).

Furthermore, it was pointed out that the lack of a clearly demarcated room for design decisions caused some of the teams to push the complexity out of their subsystems, as a way to address it under the time constraints posed by the project. Several teams independently breaking down and pushing out this complexity ---which ended on other subsystems--- led to a system architecture that lacked proper layering %
and an overall higher systemic complexity. Consequently, in LOFAR \problem{subsystem-specific design decisions also led to a suboptimal system architecture}. On top of that, \problem{there was no central architect/architectural team to address the impact of the aforementioned design decisions at component/subsystem level and to guard/drive the overall system performance}.

\subsubsection{Implementation-related problems}

Subsystems integration during Architectural Implementation in LOFAR was perceived as a challenging task that required much low-level tweaking and trouble-shooting in the initial system operations, causing delays in the project. To a large extent, the problem was in implementing the interfaces. Particularly, it was argued that \problem{subsystem interfaces were not thoroughly thought out}, and  \problem{subsystems interfaces did not  sufficiently capture the mutual expectations of the involved parties} (e.g., developers and instrument operators, hardware and software developers, etc.) Furthermore, it was pointed out that this lack of `thoroughly thought out' interfaces in the early stages of the architecting process also led to design decisions that made the project incur additional costs in different areas of the telescope operations (due, e.g., to operations that could have been automated, but ended being performed through expensive, manual labor).

When it comes to the integration of hardware and software subsystems, two further problems were identified. First, \problem{aligning the different lengths of hardware and software development cycles was difficult}. The design, prototyping and evaluation of a hardware component in LOFAR could take between three and six months, whereas a software one, which is produced under an Agile regime, had shorter cycles of two to three weeks. Therefore, the integration of hardware, firmware and software for a single subsystem was challenging as these development cycles needed to be aligned. Second, \problem{the dynamic behavior of the hardware seemed to be overlooked}.  It was pointed out that hardware is designed based on requirements focused mostly on the steady-state of the system, leaving up to the software to decide when the functions offered by it should be used, and hence leading to risky assumptions on the time-behavioral features of the hardware.

\subsection{Practices, their benefits, drawbacks and points of improvement (\textit{RQ2} and \textit{RQ3})} 

\textit{RQ2} aims to identify the practices adopted in LOFAR2.0 as a consequence of the problems identified in \textit{RQ1}. \textit{RQ3} looks into the perceived benefits and drawbacks identified during the implementation of those practices, together with possible points for improvement. In the following subsections, each practice and the problem it addresses (as depicted in Fig.~\ref{fig:problems-practices-mapping}) is described. We present the results of \textit{RQ2} and \textit{RQ3} in an integrated fashion: first we describe the practice, followed by the benefits, drawbacks, and further improvement points identified for that practice.

\subsubsection*{\textbf{B1}: Rigorous Systems Engineering process --- front-loading and subsystem requirements derived from system requirements}\label{sec:front-loading}

A more rigorous SE process is one of the most important practices adopted in LOFAR2.0, to address the problems of unknown, or unclear requirements \problemref{P1}. In this SE approach, use cases and scientific requirements are the foundation of the system-level requirements and the Operational Concept Description (OCD) document --- a system-centric description of how the system is expected to be used, considering its users, uses, and the external elements that influence its operation. Both of them are developed following a ``front-loading" strategy, which seeks to improve development performance by shifting the identification and solution of design problems to earlier phases of a product development process~\cite{thomke2000effect}. These system-level requirements and the OCD are expected to flow down to lower-level subsystems, both hardware- and software-intensive, so they can be translated, where applicable, into proper software-requirements.

\beginmrtable
		
		 \addtbrow{benefitsbg}{
		 	A more rigorous Systems Engineering process, where properly defined use cases are elicited, and interfaces are properly identified and defined \problemref{P6}, was acknowledged as key to clarifying mutual expectations (e.g., on who has to deliver which piece of software) \problemref{P7}.
	 	}
		 \addlasttbrow{1em}{benefitsbg}{benefitsfg}{Benefits}{
		 	More cooperation between groups is observed, particularly between hardware, software, and firmware-related ones. Therefore, it is fair to say that the adoption of the process is already removing a lot of the issues pertaining to the interface between these groups/disciplines.
	 	}
		 
		 \hline

\closemrtable

\vspace{-15pt}
	
\beginmrtable

		 \addtbrow{drawbacksbg}{

		 	In some cases, the information on how the subsystems (particularly software-intensive ones) should play their role within the system was lacking. This led to assumptions during the functional analysis on desired behavior, functionalities, and dynamic behavior over time. It was argued that this was caused by: (1) the resource constraints created by the multiple projects the organization is running, which, in turn,  caused the \textit{Operational Concepts} (from the OCDs) to not make it to the software teams on time; (2) system-level requirements that were either lacking or not good enough to be translated into software requirements. 
		 }
		\addlasttbrow{1em}{drawbacksbg}{drawbacksfg}{Drawbacks}{
			In LOFAR2.0, operations personnel and maintenance engineers were assigned as responsible for translating OCDs into requirements of sufficient quality for people to work with. It is perceived that this has not worked well, as that is not their job or their specialty. 
		}

\closemrtable

\vspace{-17pt}
\beginmrtable
	 	 \addtbrow{poibg}{
	 	 	 According to the study participants' experience, not every scientist is a good software engineer, and not every software engineer is able to translate scientific requirements into functional requirements on the software. Hence, in scientific software a role that serves as an interface that translates scientific use cases into requirements that engineers understand is key. Rather than delegating system-level requirements specifications to operations personnel, a role should be defined in the project to be the interface between operations staff and design team. %
	 	 }				
		\addtbrow{poibg}{ 
			Is is perceived that, for front loading, much more manpower is required from the start of the process, particularly architects and people that manage operations. 
		}
		\addlasttbrow{10em}{poibg}{poifg}{Points of Improvement}{
			More resources should be focused on producing clearer requirements, and on the definition of a high-level system architecture. %
		}
		\hline	

\closemrtable

\subsubsection*{\textbf{B2}: Hierarchical relation between architects and adoption of co-architecting/co-design}\label{sec:hierarchical-codesign}

A lesson learned from LOFAR was to avoid sequential design approaches, i.e., producing the hardware first, then the firmware, and, at the end, the software. This should help to avoid the problems of not having subsystem-level design decisions with boundaries not clearly demarcated by system-level requirements~\problemref{P3}, and the negative impact on the overall project due to local optimization decisions \problemref{P4}. Consequently, in LOFAR2.0 there is more co-architecting, where the system architect has a hierarchical relationship with the software and hardware architects; the system architect oversees both hardware- and software-intensive subsystems and components, addressing the impact of the decisions at subsystem level \problemref{P5}. The software architect, in turn, oversees the software development teams working on the various software components, which in some cases involves co-design with hardware people.

\beginmrtable
		\addlasttbrow{-1em}{benefitsbg}{benefitsfg}{Benefits}{
			
			Groups working on different subsystems are able to collaborate more closely, and much more attention is being paid to the effect of (proposed) subsystem changes to the whole system.
		}
		
		\hline
\closemrtable
\vspace{-21pt}

\beginmrtable
		
		\addtbrow{drawbacksbg}{
			Despite the adopted hierarchical co-architecting practice, 
			hardware is still frequently being designed first with little involvement of software people. Although software is meant to be flexible, not having software people involved early in the design process makes it more difficult to build consistent and high-quality components that integrate hardware and software.}	
		
		\addlasttbrow{-3em}{drawbacksbg}{drawbacksfg}{Drawbacks}{
			The SE side of the hierarchical co-architecting process must ensure the delivery of key information for the definition of a proper software architecture early in the project, including the operational requirements (as also discussed in Section \ref{sec:front-loading}) and the description of the dynamic behavior~\problemref{P9}. Defining a software architecture based on assumptions imposed by potentially missing information requires a lot of effort. On top of that, adapting the software is more expensive than having included the aforementioned elements early in the design.
		}

		\hline
\closemrtable

\vspace{-20pt}

\beginmrtable
		
		 \addtbrow{poibg}{
			Early involvement of operational teams is key in helping to describe how the system works, including the time-behavioral aspects (e.g., the dynamicity in switching signal paths). This would make way more clear what needs to be done by the software engineers, and  easier for them to start discussing interfaces with hardware people early on. Consequently, it would contribute positively to the co-design between hardware and software teams.
		}			
		\addtbrow{poibg}{
			Although software developers are able to take up architecting roles, the role of software architect at institutional level is necessary and can still be further developed. 	
		}		
		\addtbrow{poibg}{ 
			When it comes to hardware and software co-architecting, it is important to keep ---to some extent---  the flexibility of the software, as new things could be discovered while implementing it. To do so, it is important to identify the right balance between clarity and level of detail of high-level requirements, so software can remain flexible. 
		}		
		\addlasttbrow{5em}{poibg}{poifg}{Areas of improvement}{
			It was perceived that in LOFAR2.0, for the most part, hardware cycles dominate decision making not only because of their longer duration~\problemref{P8}, but also due to the background of the top management of the project clearly favoring traditional, waterfall-based approaches to system design.
			Having more people at the top of the organization with experience in Agile development approaches would probably help to shift towards more balanced hardware and software cycles.%
		}
		\hline

\closemrtable

\subsubsection*{\textbf{B3}: Splitting the system along the subsystem axis, with boundaries demarcated by ICDs}

One of the practices adopted in LOFAR2.0 was splitting up the system on a subsystem level, instead of on a hardware-firmware-software line. This new system splitting approach lead to a combination of hardware- and software-intensive subsystems, each one with a software component and data/control interface defined in an \textit{Interface Control Document} (ICD). This practice was adopted due to the problems experienced in LOFAR as a consequence of the suboptimal system architecture created by subsystem-level design decisions \problemref{P4}. Consequently, it aims for a clearer system architecture which is easier to develop in parallel by separate teams working with clearer responsibilities at subsystem level \problemref{P2}.

\beginmrtable
			\addtbrow{benefitsbg}{
				Despite the definition of ICDs still being under way in LOFAR2.0, their importance from the implementation perspective is already clear, especially for the software side: ICDs can be seen as the lifeline that connects the software to the project. 
			}
			\addlasttbrow{1em}{benefitsbg}{benefitsfg}{Benefits}{		
				This approach is perceived as positive from the software engineering perspective, as with boundaries defined by clearly defined interfaces, software-side development can be focused more on the software itself. 
			}
			\hline
\closemrtable
\vspace{-18pt}
\beginmrtable
			\addtbrow{drawbacksbg}{
				Despite the efforts to split the system at subsystem level during design, maintenance still operates on the hardware/firmware/software split. As a result, maintaining subsystems compartmentalized at subsystem levels suffers through inappropriate or suboptimal definitions of the boundaries of the involved systems.}
			\addlasttbrow{1em}{drawbacksbg}{drawbacksfg}{Drawbacks}{
				In LOFAR2.0 the ICDs are focused mainly on providing function lists, and less on defining the behavior ---initiated by such functions--- between the components involved in the described interface. This results in issues during software development~\problemref{P9}.
			}
			\hline	
\closemrtable
\vspace{-18pt}
\beginmrtable
		\addtbrow{poibg}{ 
			The firmware, an element that falls between the boundaries of hardware and software in the system, has had little involvement of software people for its design and control. It should formally be under the control of software architects.
		}
		\addtbrow{poibg}{ 
			 There is an agreement on the importance of uniformity at the interface level (i.e., ICDs), and so elements such as the monitoring points and the control points should look the same to avoid confusion and misinterpretation. 
		 }
		
		\addtbrow{poibg}{ 
			Uniformity at the implementation level should also be prioritized, so that when sub-systems are integrated, operations people will face less maintenance problems due to the diversity of languages, dependencies, coding standards, etc. 
		}
		
		\addlasttbrow{6em}{poibg}{poifg}{Areas of improvement}{
			When software is needed on one of the two sides of an ICD or if the potential client of an interface described in an ICD is software or can be software, a software engineer should be involved.
		}
		
		\hline

\closemrtable

\subsubsection*{\textbf{B4}: Early integration --- adoption of a common interface standard, and early hardware tryouts}

In LOFAR, subsystems integration was challenging and resulted in delays, particularly due to the low-level tweaking and trouble-shooting required in the initial system operations. Consequently, early integration was considered crucial for LOFAR2.0 and so development efforts were planned around integration steps. An approach that has helped drawing clear lines between subsystems and facilitated this early integration has been the adoption of a common interface standard --- in this case, the OPC-UA\footnote{OPC Unified Architecture (OPC UA), a machine to machine communication protocol for industrial automation developed by the OPC Foundation.} communication protocol. Likewise, early tryouts and debugging on hardware with prototypes are conducted aiming at starting their integration with their software counterparts early in the development process \problemref{P7}. Given that most hardware in LOFAR2.0 interacts with software through a microcontroller, microcontrollers that are not yet connected to the hardware but behave according to their ICD have worked as the means for prototyping (particularly into the control system), and hence, for early integration.

\beginmrtable

	\addlasttbrow{-1em}{benefitsbg}{benefitsfg}{Benefits}{		
		The `early debugging' and `early tryouts' practices, when clear interfaces have been defined between hardware and software (e.g., through ICDs), help the software-side to flag, early in the project, when the hardware is not working well.
	}

\closemrtable

\vspace{-18pt}

\beginmrtable
		
		\addlasttbrow{-1em}{drawbacksbg}{drawbacksfg}{Drawbacks}{ 			
			Even though the adoption of interface standards and approaches for early hardware tryouts have proven to be helpful for early integration, the seeming lack of coordination between hardware development cycles and software still makes it difficult. Such lack of coordination, in turn, is attributed to the long cycles of hardware, including manufacturing processes.
		}

		\hline
\closemrtable

\vspace{-18pt}
\beginmrtable

	\addtbrow{poibg}{ 
		A good practice to promote early integration would be to start the development of ICDs early in the project --- including what is known about them at the moment, and treating them as a living document, instead of trying to make them perfect and complete from the beginning. With these ICDs, software engineers would not only be able to start building their components early on, but also seeing them as a communication tool to get further explanations from the people working at the other end.
	}
	\addlasttbrow{1em}{poibg}{poifg}{Areas of improvement}{
	    Sub-system emulators, derived from properly defined ICDs are considered as a key element for early integration, as they facilitate the development and testing ---particularly software--- of subsystems that are in a different development path. For this reason, every new component should have an emulator part that is easy to be configured. However, it is also important for the project managers to acknowledge the value of these emulators, or `digital twins', so the budget for their development is assigned early in the project.
	}

		\hline
		
\closemrtable

\subsection{Problems from LOFAR not addressed yet or still explored in LOFAR2.0}

There are two identified problems for which no best practices were identified by the participants of this study as shown in Fig.~\ref{fig:problems-practices-mapping}.
First, the problem of aligning the development cycles for the integration of hardware, firmware, and software in a subsystem \problemref{P8}, is something still under consideration (at the moment of our data collection) by LOFAR2.0 system architects, through looking at experiences in other domains.
In addition, and due to the Systems Engineering process adopted,  the integration of the front-loading approach on System Requirements, with the Agile Scrum methodology used for software development has also emerged as a challenge in LOFAR2.0. An approach to tackle this problem, is seemingly taking shape at the moment in LOFAR2.0.
In this approach, software-related requirements are taken as customer needs, from which a sequence of MVPs (Minimum Viable Products)  are defined, each one with increasing business value. Consequently, the scrum teams work towards each MVP, adjusting its scope (e.g., prioritizing functionality over technical debt) on each sprint. 

The other unaddressed problem is the lack of dynamic behavior-related information in the requirements \problemref{P9}. %
More specifically, it was pointed out that time-behavioral information such as \textit{what should happen} to get from one steady-state to the other, \textit{how often}, or \textit{what happens if a failure takes place in-between}, is not sufficiently specified by the requirements. In LOFAR2.0 some software teams needed to figure this out on their own, hence leading to very similar risky assumptions to  the ones described in problem \problemref{P9} for LOFAR. One of the participants illustrated this with an analogy with a car design, whose requirements are limited to features such as ABS, automatic lane control, and ability to drive at 100\,km/h on a highway. With no further time-behavioral information (e.g., how the top speed should be reached), the car designers could omit key features like a clutch and a gearbox to actually reach the cruise control state.

	\section{Discussion}
	\label{sec:discussion}

		This study analyzed the practices for harmonizing system- and software-level architecting activities adopted by practitioners in a set of long-term projects developing a large and complex scientific instrument, looking also into the problems that motivated them and their perceived pros and cons. In this section we elaborate on the interpretation and implications of the results for practitioners and researchers in the domain of SoS architecting in general. 

A previous study by Cadavid et al.~\cite{cadavid2020survey} already found that mutual assumptions at system- and software-level are a major issue during SoS architecting. In this study, we confirmed this insight, discovering that, \lesson{in large scale systems, assumptions on the dynamic behavioral aspects of software-intensive subsystems such as the ones related to time are, indirectly, a major cause of budget, integration and operation-related problems}. For instance, time-behavioral assumptions seem to lead to software architecture and design decisions that eventually require fairly deep and expensive rework, accruing architectural debt in the process. The fact that this rework potentially requires more resources than the ones required for reaching an agreement on this matter early on, highlight \lesson{the importance of not overestimating the flexibility of software in this kind of systems}. Overall, the negative impact of insufficient emphasis on the dynamic (behavioral) elements of the sub-systems,  and the communication challenges in multi-disciplinary engineering teams (cf.~\cite{cadavid2020survey,fairley2019systems}) highlight the importance of proper approaches for specifying said dynamic behavior.

According to the study results, the origin of these assumptions lies ultimately at the system-level requirements. In LOFAR, there were problems of missing or unclear system-level requirements as, back then, the system was a unique instrument concept, with unclear features yet to be discovered\footnote{To be more precise, LOFAR at least at its inception was a technological project taking advantage of a \textit{processing window of opportunity}. This is evident in the identified problems by this study, where no system-level scientific requirements were brought up by the participants.}~\cite{Bregman:1999,Bregman:2000,Bregman:2004}. %
Although in LOFAR2.0  the adoption of a more rigorous SE process, with a \textit{front-loading} approach on requirements was aimed to address this, requirements-related problems persisted, but this time due to the domain knowledge required by software engineers to understand scientific use cases, a common issue also reported in~\cite{cadavid2020survey}. Previous studies~\cite{muscarella2020systems} highlighted the importance of interdisciplinary communication as a critical component of the process to avoid these kinds of problems. However, in the particular context of this study, it was pointed out that people talking with each other does not necessarily mean that they will understand each other. Therefore, \lesson{effective cooperation between system- and software-related disciplines is not only about communication, but mainly about the inclusion of an actual role that serves as an interface between the two disciplines and the rest of the project stakeholders}.

Moving from analysis to synthesis, the practice of splitting the system along the subsystems axis, and the use of ICDs for subsystem demarcation was a significant improvement during the LOFAR2.0 system design. It is worth noting that the use of ICDs for the separation of concerns in the design of loosely coupled subsystems has been extensively discussed in empirical research literature~\cite{champagne_will_2016,blyler_interface_2004,wheatcraft20109}, which identifies them as a key element for improving the interfacing between SE and SWE~\cite{muscarella2020systems}. The results of this exploratory case study, however, suggest that \lesson{the \emph{uniformity} of ICDs across the system is key, particularly in large-scale systems}. On systems like LOFAR2.0, \lesson{monitoring and control points, in addition to the I/O-related ones~\cite{muscarella2020systems}, are essential elements to be considered and uniformly described in their ICDs}.

When it comes to the practice of hierarchical co-architecting and co-design, it is worth noting that the drawbacks and points for improvement that emerged from the discussions were more oriented towards hardware-software co-architecting rather than system-subsystem co-architecting. One of the findings from these discussions, which could be attributed to the uniqueness of the scientific systems' hardware, is the importance of the involvement of software engineers early in hardware design. That is to say, regardless of the existence of design artifacts such as  the ICDs to avoid dependencies during the development of interrelated hardware and software, \lesson{proper software involvement early in hardware design still seems to be key for building consistent, high-quality hardware/software frameworks}. Once again, and for the sake of the overall quality of large-scale systems' implementation, this is a call to avoid over-reliance on the flexibility of software as a fix-all solution.

Given the reported experiences in LOFAR2.0 and similar scientific instruments~\cite{tanci2016software}, \lesson{adopting communication standards for the subsystems interfaces like UPC-UA seems to be a promising practice to improve not only early integration in the project, but also the unification of ICDs}; this topic is worth further exploration on its own merit. In the case study, simulators or \textit{digital twins} also arose as valuable tools for early integration and testing, which is also consistent with the best practices identified in~\cite{muscarella2020systems}. \lesson{The suggested point of improvement of using ICDs in order to derive digital twins, also sheds light on the importance of automating the generation of these artifacts}~\cite{haag2019automated,campos2019automatic}. Furthermore, the positive early integration experiences by using the microcontrollers that are meant to be the front-end of the hardware, but are not yet connected to it, suggest \lesson{the importance of further research and development on microcontroller-based digital twins, as done in other domains}~\cite{hinchy2019using}.

Overall, we believe that these findings can be useful beyond the radio astronomy domain and systems like LOFAR. Similar problems have been explored in other domains featuring systems with similar SoS-characteristics, i.e., complex, large scale systems, despite not being self-identified as SoS --- a common phenomenon in the literature, as reported in~\cite{cadavid2019architecting}. These problems include  requirements-related ones in \textit{Smart Cities}~\cite{assiri2020software}; interface management issues in \textit{Intelligent Transport Systems}~\cite{bhardwaj2019designing}, \textit{Avionics}~\cite{louadah_data_2020}, and \textit{FPSOs}\footnote{Floating production storage and offloading}~\cite{yasseri2019interface}; and life-cycles harmonization-related in \textit{Automotive} systems~\cite{kasauli2020agile}. Further research to collect evidence on the cross-domain nature of our findings is the subject of future work in any case.

	\section{Lessons Learned}	
	\label{sec:lessons}
    
    On a relevant note, and despite not being an explicitly identified goal of this work, we would like to briefly reflect on the methods and techniques we chose for collecting qualitative data in this study during the COVID-19 pandemic, and the lessons we learned in this process.
    
    Our choice of an asynchronous approach for interviews and focus groups is the result of the necessity to conduct the study under partial or complete lockdown conditions, while also attempting to minimize the effort of arranging, conducting, and transcribing online interviews and group discussions.
    Given the fact that the daily activities of both the practitioners and the researchers of this study were clearly affected by the need to adjust to working from home, in addition to the overall disruption and psychological pressure induced by the pandemic, we feel that this choice is justified: it allowed both practitioners and researchers to contribute based on their own schedule and availability.
     
    The email-based interviews worked particularly well, allowing: interviewees to articulate their thoughts and opinions in a more organized manner; more time for us to analyze their responses in more depth and point out items for follow-up questions; and both sides to get familiar with the terms being used from the different domains.
    Especially the last part would have been very difficult to manage if the interviews had taken place in person or through a teleconferencing system.
    
    Running a text-based, asynchronous virtual focus group, on the other hand, was a mixed bag of experiences.
    Some discussion points picked up a lot of responses and reactions, but most of them had only a few or no reactions from the participants, despite regular reminders and explicitly identifying participants that could provide specific insights to the discussed point based on their profile.
    Furthermore, as the Slack channel started getting ``older'' the amount of interactions was continually diminishing to the point of stagnation during what would under normal conditions be the summer holidays period.
    This essentially forced our hand to arrange for an online Zoom call with all active channel participants to go over already brought up points that needed further clarifications, and discuss some points that had not been introduced in the channel yet.
    Participation in the Zoom call was high, and the amount of data collected during the two hour call was basically equivalent, if not greater, to what we managed to collect in running the channel for around two months.
    Even in this case, however, running the Slack channel for as long as we did, provided us with space and time to get familiar with the domain and the case itself, which definitely contributed to a more productive Zoom call with the participants in the last stage.
    
    Under this light, we cannot in principle advise to solely use a text-based virtual focus group as the primary means for data collection from a practitioner group. However, we do recommend it instead as a side-channel of communication with the participants, in preparation for online video meetings. 
    At the same time, and with the caveat that we have no way or intention for controlling for it, we feel that overall the online mode of data collection did not necessarily result into a lesser amount or lower quality of data being collected in comparison to doing the same study in-person.
    The fact that both practitioners and researchers had to adjust to exclusively online communications with their colleagues during this period anyway may have contributed to this.

	\section{Limitations and threats to validity}\label{sec:ttv}	
	
In this section we present potential threats to validity of this study, and our actions for their mitigation. We use Runeson and H\"ost~\cite{runeson2009guidelines} as a guide for this purpose. It is worth noting that this study is not subject to internal validity threats, as it does not investigate causality.

\subsubsection{Construct validity}

Construct validity refers to the degree to which the operational measures, in this case the interviews and the virtual focus group, reflect what the researchers have in mind and what is investigated according to the research questions. To mitigate this, the seed questions of the interviews were reviewed from two complementary viewpoints: the domain-specific perspective of one of the co-authors who is a member of ASTRON, and the empirical research perspective of the senior researcher involved in the study. The VFG environment and seed questions, on the other hand, were not only reviewed by the same two researchers, but also piloted by a group of engineers from the researchers' institutions that were not consequently involved in this study. This provided key feedback on the VFG seed questions, and on how to deal with the synchronous and asynchronous dynamics of using Slack for the VFG.

\subsubsection{Reliability}
Reliability refers to the extent which the data collection and analysis are dependent on the specific researchers. To mitigate potential threats in this respect, the guidelines of Runeson and H\"ost~\cite{runeson2009guidelines} were used as a general framework for the research, and commonly cited guidelines for conducting interviews and focus groups in a virtual setting were followed for more specific research practices~\cite{bowden2015interviewing,stewart2005researching,murray1997using,moloney2003using}.
To mitigate research bias in data analysis, three out of the four authors  were involved in the data collection process, and came to an agreement through consensus building about the interpretations drawn from the analysis of the collected data. This process can be independently verified and traced back to the original quotes via the available replication package\footnote{\url{https://doi.org/10.6084/m9.figshare.13332659.v1}}.

\subsubsection{External validity}
External validity is concerned with the degree to which the findings can be generalized from the sample to the population. Our findings on harmonization practices for long-running large-scale scientific instrument projects are based on a very specific setting, which limits their generalizability. More work is definitely required to further explore the generalizability of our findings in other domains, e.g., by means of confirmatory case studies. However, as discussed further in Section~\ref{sec:discussion}, we believe that the characteristics of the case subjects (LOFAR and LOFAR2.0 projects), namely their size, complexity, and  physical distribution in combination with managerial independence, make our findings relevant for a wider class of SoS with similar characteristics such as smart cities and smart factories in Industry 4.0.

	\section{Conclusions and future work}
	\label{sec:conclusion}

		The architecting process of a large-scale System of Systems involves a myriad of challenges, one of the most important being the problematic interplay between the disciplines involved at different levels. For instance, it has been reported that the gaps and mismatches between the architecting processes at both system- and software-level are often linked to major integration and operational problems, not only in SoS, but in engineered systems in general. Given the scarcity of guidelines to improve these gaps and mismatches in SoS, this study explores the system- and software-level architecting harmonization practices that have been discovered by practitioners involved %
in a long-running SoS project. For this purpose, we conducted an exploratory case study at ASTRON, the Netherlands Institute for Radio Astronomy, where the LOFAR radio telescope project, and its follow up, LOFAR2.0 ---which together add more than 20 years of architecting experience--- were analyzed.
 
The results of this study suggest that systems engineering processes must give special attention to the way scientific requirements are translated into system requirements, so said system requirements can be properly translated into software requirements.  In particular, we found that system-level requirements should be clear and detailed enough, but at a level where software keeps its flexibility. At the same time, this flexibility of software should not be overestimated since it often results into increased development costs down the line.   

Along the same lines, it is key to provide elements that make clear the role of the software subsystems in the grander scheme of the whole SoS, that is to say, going beyond the functional elements, and providing in addition at least time-behavioral ones. This requires ---especially in scientific instruments--- early involvement of operations people in the architecting process, and a role specialized on the translation of scientific requirements. 

We also found that uniformity is a key element for integration-oriented artifacts such as ICDs in such large-scale projects, and that such uniform elements should include control and monitoring points, elements that are seemingly often omitted when produced exclusively by hardware engineers. Furthermore, the study results highlight again the importance of not  over-relying on the flexibility of the software when designing hardware. Our findings also suggest that facilitating an early involvement of software architects in hardware design not only improves the quality of the hardware-software frameworks, but significantly reduces the cost of software adaptations. Last, but not least, emerging communication standards and microcontroller-based digital twins seem to be promising approaches for the aforementioned interface unification and early testing/integration purposes, respectively.

As a future work, we plan to further explore the generalizability of our findings by means of confirmatory case studies across domains. Furthermore, we also plan to build onto these findings, particularly by devising architecting artifacts that would improve uniformity in hardware-software interface specifications, while enabling the process of deriving digital twins from them.

	\section*{Acknowledgments}
	This work was supported by ITEA3 and RVO under grant agreement No. 17038 VISDOM (\url{https://visdom-project.github.io/website/}). 
	
	The authors would like to thank the participants of the case study for volunteering their invaluable time in helping us with conducting this research during a difficult period in their and our lives. We would also like to thank Ren\'{e} Kaptijn for facilitating the process and \'{A}gnes Mika for her feedback on an early version of this manuscript. We are grateful to the anonymous reviewers for their comments on this paper.
	
	\medskip
	
	\bibliographystyle{unsrt}
	\bibliography{refs,dcs,sos-patt-archs,sec-studies,research-method,icd-refs,related-non-sos}
	
\end{document}